\documentclass[%
reprint,
superscriptaddress,
showpacs,
 amsmath,amssymb,
 aps,
 prl,
]{revtex4-1}

\usepackage{graphicx}
\usepackage{dcolumn}
\usepackage{bm}
\usepackage{amsmath}
\usepackage{autobreak}
\usepackage{hyperref}
\usepackage[mathlines]{lineno}
\usepackage{natbib}
\usepackage{multirow}

\begin{document}

\preprint{APS/123-QED}

\title{Constraints on Spin-Independent Nucleus Scattering with sub-GeV Weakly Interacting Massive Particle Dark Matter from the CDEX-1B Experiment at the China Jin-Ping Laboratory}

\affiliation{Key Laboratory of Particle and Radiation Imaging (Ministry of Education) and Department of Engineering Physics, Tsinghua University, Beijing 100084}
\affiliation{Institute of Physics, Academia Sinica, Taipei 11529}

\affiliation{Department of Physics, Dokuz Eyl\"{u}l University, \.{I}zmir 35160}
\affiliation{Department of Physics, Tsinghua University, Beijing 100084}
\affiliation{NUCTECH Company, Beijing 100084}
\affiliation{YaLong River Hydropower Development Company, Chengdu 610051}
\affiliation{College of Nuclear Science and Technology, Beijing Normal University, Beijing 100875}
\affiliation{College of Physical Science and Technology, Sichuan University, Chengdu 610065}
\affiliation{School of Physics, Peking University, Beijing 100871}
\affiliation{Department of Nuclear Physics, China Institute of Atomic Energy, Beijing 102413}
\affiliation{School of Physics, Nankai University, Tianjin 300071}
\affiliation{Department of Physics, Banaras Hindu University, Varanasi 221005}
\affiliation{Department of Physics, Beijing Normal University, Beijing 100875}
\author{Z.~Z.~Liu}
\affiliation{Key Laboratory of Particle and Radiation Imaging (Ministry of Education) and Department of Engineering Physics, Tsinghua University, Beijing 100084}
\author{Q. Yue}\altaffiliation [Corresponding author: ]{yueq@mail.tsinghua.edu.cn}
\affiliation{Key Laboratory of Particle and Radiation Imaging (Ministry of Education) and Department of Engineering Physics, Tsinghua University, Beijing 100084}
\author{L.~T.~Yang}\altaffiliation [Corresponding author: ]{yanglt@mail.tsinghua.edu.cn}
\affiliation{Key Laboratory of Particle and Radiation Imaging (Ministry of Education) and Department of Engineering Physics, Tsinghua University, Beijing 100084}
\author{K.~J.~Kang}
\affiliation{Key Laboratory of Particle and Radiation Imaging (Ministry of Education) and Department of Engineering Physics, Tsinghua University, Beijing 100084}
\author{Y.~J.~Li}
\affiliation{Key Laboratory of Particle and Radiation Imaging (Ministry of Education) and Department of Engineering Physics, Tsinghua University, Beijing 100084}
\author{H.~T.~Wong}
\altaffiliation{Participating as a member of TEXONO Collaboration}
\affiliation{Institute of Physics, Academia Sinica, Taipei 11529}
\author{M. Agartioglu}
\altaffiliation{Participating as a member of TEXONO Collaboration}
\affiliation{Institute of Physics, Academia Sinica, Taipei 11529}
\affiliation{Department of Physics, Dokuz Eyl\"{u}l University, \.{I}zmir 35160}
\author{H.~P.~An}
\affiliation{Key Laboratory of Particle and Radiation Imaging (Ministry of Education) and Department of Engineering Physics, Tsinghua University, Beijing 100084}
\affiliation{Department of Physics, Tsinghua University, Beijing 100084}
\author{J.~P.~Chang}
\affiliation{NUCTECH Company, Beijing 100084}
\author{J.~H.~Chen}
\altaffiliation{Participating as a member of TEXONO Collaboration}
\affiliation{Institute of Physics, Academia Sinica, Taipei 11529}
\author{Y.~H.~Chen}
\affiliation{YaLong River Hydropower Development Company, Chengdu 610051}
\author{J.~P.~Cheng}
\affiliation{Key Laboratory of Particle and Radiation Imaging (Ministry of Education) and Department of Engineering Physics, Tsinghua University, Beijing 100084}
\affiliation{College of Nuclear Science and Technology, Beijing Normal University, Beijing 100875}
\author{Z.~Deng}
\affiliation{Key Laboratory of Particle and Radiation Imaging (Ministry of Education) and Department of Engineering Physics, Tsinghua University, Beijing 100084}
\author{Q.~Du}
\affiliation{College of Physical Science and Technology, Sichuan University, Chengdu 610065}
\author{H.~Gong}
\affiliation{Key Laboratory of Particle and Radiation Imaging (Ministry of Education) and Department of Engineering Physics, Tsinghua University, Beijing 100084}
\author{X.~Y.~Guo}
\affiliation{YaLong River Hydropower Development Company, Chengdu 610051}
\author{Q.~J.~Guo}
\affiliation{School of Physics, Peking University, Beijing 100871}
\author{L. He}
\affiliation{NUCTECH Company, Beijing 100084}
\author{S.~M.~He}
\affiliation{YaLong River Hydropower Development Company, Chengdu 610051}
\author{J.~W.~Hu}
\affiliation{Key Laboratory of Particle and Radiation Imaging (Ministry of Education) and Department of Engineering Physics, Tsinghua University, Beijing 100084}
\author{Q.~D.~Hu}
\affiliation{Key Laboratory of Particle and Radiation Imaging (Ministry of Education) and Department of Engineering Physics, Tsinghua University, Beijing 100084}
\author{H.~X.~Huang}
\affiliation{Department of Nuclear Physics, China Institute of Atomic Energy, Beijing 102413}
\author{L.~P.~Jia}
\affiliation{Key Laboratory of Particle and Radiation Imaging (Ministry of Education) and Department of Engineering Physics, Tsinghua University, Beijing 100084}
\author{H. Jiang}
\affiliation{Key Laboratory of Particle and Radiation Imaging (Ministry of Education) and Department of Engineering Physics, Tsinghua University, Beijing 100084}
\author{H.~B.~Li}
\altaffiliation{Participating as a member of TEXONO Collaboration}
\affiliation{Institute of Physics, Academia Sinica, Taipei 11529}
\author{H. Li}
\affiliation{NUCTECH Company, Beijing 100084}
\author{J.~M.~Li}
\affiliation{Key Laboratory of Particle and Radiation Imaging (Ministry of Education) and Department of Engineering Physics, Tsinghua University, Beijing 100084}
\author{J.~Li}
\affiliation{Key Laboratory of Particle and Radiation Imaging (Ministry of Education) and Department of Engineering Physics, Tsinghua University, Beijing 100084}
\author{X.~Li}
\affiliation{Department of Nuclear Physics, China Institute of Atomic Energy, Beijing 102413}
\author{X.~Q.~Li}
\affiliation{School of Physics, Nankai University, Tianjin 300071}
\author{Y.~L.~Li}
\affiliation{Key Laboratory of Particle and Radiation Imaging (Ministry of Education) and Department of Engineering Physics, Tsinghua University, Beijing 100084}
\author {B. Liao}
\affiliation{College of Nuclear Science and Technology, Beijing Normal University, Beijing 100875}
\author{F.~K.~Lin}
\altaffiliation{Participating as a member of TEXONO Collaboration}
\affiliation{Institute of Physics, Academia Sinica, Taipei 11529}
\author{S.~T.~Lin}
\affiliation{College of Physical Science and Technology, Sichuan University, Chengdu 610065}
\author{S.~K.~Liu}
\affiliation{College of Physical Science and Technology, Sichuan University, Chengdu 610065}
\author {Y.~D.~Liu}
\affiliation{College of Nuclear Science and Technology, Beijing Normal University, Beijing 100875}
\author {Y.~Y.~Liu}
\affiliation{College of Nuclear Science and Technology, Beijing Normal University, Beijing 100875}
\author{H.~Ma}\altaffiliation [Corresponding author: ]{mahao@mail.tsinghua.edu.cn}
\affiliation{Key Laboratory of Particle and Radiation Imaging (Ministry of Education) and Department of Engineering Physics, Tsinghua University, Beijing 100084}
\author{J.~L.~Ma}
\affiliation{Key Laboratory of Particle and Radiation Imaging (Ministry of Education) and Department of Engineering Physics, Tsinghua University, Beijing 100084}
\affiliation{Department of Physics, Tsinghua University, Beijing 100084}
\author{Y.~C.~Mao}
\affiliation{School of Physics, Peking University, Beijing 100871}
\author{J.~H.~Ning}
\affiliation{YaLong River Hydropower Development Company, Chengdu 610051}
\author{H.~Pan}
\affiliation{NUCTECH Company, Beijing 100084}
\author{N.~C.~Qi}
\affiliation{YaLong River Hydropower Development Company, Chengdu 610051}
\author{J.~Ren}
\affiliation{Department of Nuclear Physics, China Institute of Atomic Energy, Beijing 102413}
\author{X.~C.~Ruan}
\affiliation{Department of Nuclear Physics, China Institute of Atomic Energy, Beijing 102413}
\author{V.~Sharma}
\altaffiliation{Participating as a member of TEXONO Collaboration}
\affiliation{Institute of Physics, Academia Sinica, Taipei 11529}
\affiliation{Department of Physics, Banaras Hindu University, Varanasi 221005}
\author{Z.~She}
\affiliation{Key Laboratory of Particle and Radiation Imaging (Ministry of Education) and Department of Engineering Physics, Tsinghua University, Beijing 100084}
\author{L.~Singh}
\altaffiliation{Participating as a member of TEXONO Collaboration}
\affiliation{Institute of Physics, Academia Sinica, Taipei 11529}
\affiliation{Department of Physics, Banaras Hindu University, Varanasi 221005}
\author{M.~K.~Singh}
\altaffiliation{Participating as a member of TEXONO Collaboration}
\affiliation{Institute of Physics, Academia Sinica, Taipei 11529}
\affiliation{Department of Physics, Banaras Hindu University, Varanasi 221005}

\author {T.~X.~Sun}
\affiliation{College of Nuclear Science and Technology, Beijing Normal University, Beijing 100875}

\author{C.~J.~Tang}
\affiliation{College of Physical Science and Technology, Sichuan University, Chengdu 610065}
\author{W.~Y.~Tang}
\affiliation{Key Laboratory of Particle and Radiation Imaging (Ministry of Education) and Department of Engineering Physics, Tsinghua University, Beijing 100084}
\author{Y.~Tian}
\affiliation{Key Laboratory of Particle and Radiation Imaging (Ministry of Education) and Department of Engineering Physics, Tsinghua University, Beijing 100084}

\author {G.~F.~Wang}
\affiliation{College of Nuclear Science and Technology, Beijing Normal University, Beijing 100875}

\author{L.~Wang}
\affiliation{Department of Physics, Beijing Normal University, Beijing 100875}
\author{Q.~Wang}
\affiliation{Key Laboratory of Particle and Radiation Imaging (Ministry of Education) and Department of Engineering Physics, Tsinghua University, Beijing 100084}
\affiliation{Department of Physics, Tsinghua University, Beijing 100084}
\author{Y.~Wang}
\affiliation{Key Laboratory of Particle and Radiation Imaging (Ministry of Education) and Department of Engineering Physics, Tsinghua University, Beijing 100084}
\affiliation{Department of Physics, Tsinghua University, Beijing 100084}
\author{Y.~X.~Wang}
\affiliation{School of Physics, Peking University, Beijing 100871}
\author{S.~Y.~Wu}
\affiliation{YaLong River Hydropower Development Company, Chengdu 610051}
\author{Y.~C.~Wu}
\affiliation{Key Laboratory of Particle and Radiation Imaging (Ministry of Education) and Department of Engineering Physics, Tsinghua University, Beijing 100084}
\author{H.~Y.~Xing}
\affiliation{College of Physical Science and Technology, Sichuan University, Chengdu 610065}
\author{Y.~Xu}
\affiliation{School of Physics, Nankai University, Tianjin 300071}
\author{T.~Xue}
\affiliation{Key Laboratory of Particle and Radiation Imaging (Ministry of Education) and Department of Engineering Physics, Tsinghua University, Beijing 100084}

\author{N.~Yi}
\affiliation{Key Laboratory of Particle and Radiation Imaging (Ministry of Education) and Department of Engineering Physics, Tsinghua University, Beijing 100084}
\author{C.~X.~Yu}
\affiliation{School of Physics, Nankai University, Tianjin 300071}
\author{H.~J.~Yu}
\affiliation{NUCTECH Company, Beijing 100084}
\author{J.~F.~Yue}
\affiliation{YaLong River Hydropower Development Company, Chengdu 610051}
\author{M.~Zeng}
\affiliation{Key Laboratory of Particle and Radiation Imaging (Ministry of Education) and Department of Engineering Physics, Tsinghua University, Beijing 100084}
\author{Z.~Zeng}
\affiliation{Key Laboratory of Particle and Radiation Imaging (Ministry of Education) and Department of Engineering Physics, Tsinghua University, Beijing 100084}

\author {F.~S.~Zhang}
\affiliation{College of Nuclear Science and Technology, Beijing Normal University, Beijing 100875}

\author{M.~G.~Zhao}
\affiliation{School of Physics, Nankai University, Tianjin 300071}
\author{J.~F.~Zhou}
\affiliation{YaLong River Hydropower Development Company, Chengdu 610051}
\author{Z.~Y.~Zhou}
\affiliation{Department of Nuclear Physics, China Institute of Atomic Energy, Beijing 102413}
\author{J.~J.~Zhu}
\affiliation{College of Physical Science and Technology, Sichuan University, Chengdu 610065}

\collaboration{CDEX Collaboration}
\noaffiliation

\date{\today}

\begin{abstract}We report results on the searches of weakly interacting massive particles (WIMPs) with sub-GeV masses ($m_{\chi}$) via WIMP-nucleus spin-independent scattering with Migdal effect incorporated. Analysis on  time-integrated (TI) and annual modulation (AM) effects on CDEX-1B data are performed, with 737.1 kg$\cdot$day exposure and 160 eVee threshold for TI analysis, and 1107.5 kg$\cdot$day exposure and 250 eVee threshold for AM analysis. The sensitive windows in $m_{\chi}$ are expanded by an order of magnitude to lower DM masses with Migdal effect incorporated. New limits on $\sigma_{\chi N}^{\rm SI}$ at 90\% confidence level are derived as $2\times$10$^{-32}\sim7\times$10$^{-35}$ $\rm cm^2$ for TI analysis at $m_{\chi}\sim$ 50$-$180 MeV/$c^2$, and $3\times$10$^{-32}\sim9\times$10$^{-38}$ $\rm cm^2$ for AM analysis at $m_{\chi}\sim$75 MeV/$c^2-$3.0 GeV/$c^2$.
\begin{description}
\item[PACS numbers]{95.35.+d,
29.40.-n,
98.70.Vc}
\end{description}
\end{abstract}

\maketitle


\emph{Introduction.} Weakly interacting massive particles (WIMPs, denoted as $\chi$) are the most popular candidates of dark matter, the searches of which are of intense experimental interest~\cite{pdg2018}. Direct detection experiments such as XENON~\cite{xenon1t}, LUX~\cite{lux}, PandaX~\cite{PandaX}, SuperCDMS~\cite{cdmslite}, DarkSide~\cite{darkside}, CDEX~\cite{cdex0,cdex1,cdex12014,cdex12016,cdex1b2018,cdex102018} are based on WIMP-nucleus ($\chi$-$N$) elastic scatterings through spin-independent (SI) and spin-dependent interactions. However, the nuclear recoil energy and hence the experimental observable rapidly diminishes with decreasing $m_{\chi}$. Detectors with low threshold have to be used to study these light WIMPs. At the lowest achieved threshold of 30.1 eV in nuclear recoil energy, the CRESST~\cite{cresst} experiment extends the low reach of $m_{\chi}$ to 160 MeV/$c^2$, using the conventional $\chi$-$N$ scattering detection channel.

It has been pointed out that finite amount of electrons or photons are produced in $\chi$-$N$ inelastic scattering~\cite{migdal0,migdaleffect,migdaleffectprl,bremsstrahlung}. Two of the mechanisms that produce electro-magnetic final states in $\chi$-$N$ scatterings are Migdal effect~\cite{migdaleffect,migdaleffectprl} and bremsstrahlung emissions~\cite{bremsstrahlung}. The observable signals due to electron recoils or gamma rays are much larger than those of nuclear recoils at $m_{\chi}$ less than a few GeV. Taking these two effects into account, the lower reach of $m_{\chi}$ in direct detection experiments can be substantially extended to domains far below 1 GeV/$c^2$. 

$P$-type point contact germanium (PPCGe) detectors have been adopted by CDEX~\cite{cdex0,cdex1,cdex12014,cdex12016,cdex1b2018,cdex102018} in light WIMP searches, exploiting their good energy resolution and ultralow energy threshold. Located in the China Jinping Underground Laboratory (CJPL)~\cite{cjpl}, CDEX-1B experiment~\cite{cdex1b2018} has for its target a single-element PPCGe detector cooled by a cold finger, with an active mass of 939 g. A NaI(Tl) detector is used as active shielding to veto the gamma-ray induced background events. The detector has been under stable data taking conditions since March 27, 2014 and limits on $\chi$-$N$ SI-scattering down to $m_{\chi}$$\sim$2 GeV/$c^2$ are derived at an energy threshold of 160 eVee (``eVee" represents the electron equivalent energy) with an exposure of 737.1 kg-day~\cite{cdex1b2018}. The detector CDEX-1B has been working stably and the data obtained have good time stability, so the data are also used for annual modulation analysis~\cite{C1B-AM-paper-underreview}. In this letter, taking Migdal effect into account, the CDEX-1B data is reanalyzed to derive new limits on WIMP-nucleon SI-interactions, cross section of which denoted by $\sigma_{\chi N}^{\rm SI}$.

\emph{Migdal effect.}— The conventional and simplified treatment of $\chi$-$N$ scattering is that all the kinetic energy is transferred from $\chi$ to nuclear recoil via elastic scattering. Complexities arise in real physical systems, since the target nuclei in detectors, being part of the atoms, are coupled also to the electrons. There is finite probability that high-energy electrons are ejected via inelastic $\chi$-N scattering processes. The electrons do not follow the motion of the nuclei such that the electrons of the target atom will be excited or ionized. The process, called Migdal effect~\cite{migdalorigin,migdalorigin0,migdalorigin2} was recently studied in the context of WIMP detection via $\chi$-$N$ interactions~\cite{migdaleffect,migdaleffectprl}. According to Ref.~\cite{migdaleffect}, the ionization of a single electron is the dominant effect, whereas multielectron ionization and excitation as well as single-electron excitation can be neglected. Accordingly, only single electron ionization is considered in this analysis.

After an electron is ejected via ionization, the ionized atom will deexcite and emit new electrons or photons, whose total energy is the binding energy, denoted as $E_{nl}$. The total electronic energy, distinctive from the nuclear recoil energy, is given by $E_{EM}=E_{e}+E_{nl}$, where $E_{e}$ is the kinetic energy of electron after ionization, while the cross section is given by 
\begin{equation}
\begin{aligned}
\frac{d^2\sigma}{dE_RdE_{EM}} = \frac{d\sigma}{dE_R} \frac{1}{2\pi }\sum_{n,l} \frac{d}{dE_{EM}}p_{q_e}^c(nl\to (E_{EM}-E_{nl})),
\end{aligned}
\end{equation}
where $E_{R}$ is the nuclear recoil energy, $p_{q_e}^c$ is the ionization probability, $q_{e}$ is equal to $m_{e}q_{A}/m_{A}$ , $m_{A}$ is the atomic mass approximated to target nucleus mass, $q_{A}$ is equal to $\sqrt{2m_{A}E_{R}}$, $n$ and $l$ are the principal quantum number and orbital quantum number, respectively~\cite{migdaleffect}.  

The maximum electronic energy $E_{EM,max}$ is equal to $1/2\mu_{N}v^2$, $\mu_N$ is the reduced mass between $\chi$ and target nucleus, $v$ is the relative velocity between $\chi$ and the target nucleus, while the maximum energy of nuclear recoil $E_{R,max}$ is equal to $2\mu_{N}^2v^2/m_N$~\cite{migdaleffect,migdaleffectprl}. If $m_\chi \ll m_{N}$, then $\mu_N \sim m_\chi$, such that $E_{EM,max}\gg E_{R,max}$. For example, while $m_\chi = 50$ MeV/${c}^2$ and $m_N=67.66$ GeV/${c}^2$ (the nucleus mass of Ge), for $v_{max}$=776 km/s, the resulting $E_{EM,max}\approx 167 $ eV and $E_{R,max}\approx 1$ eV.
 
The $\chi$-$N$ event rates due to Migdal effect can be expressed as:
\begin{equation}
\begin{aligned}
\frac{d^2R}{dE_{EM} dE_{R}} = 
N_{T} \frac {\rho_\chi}{m_\chi} \int d^3 \vec{v} vf_v(\vec{v}+\vec{v}_E)
\frac {d^2\sigma}{dE_{EM}dE_R},
\end{aligned}
\end{equation}
where $N_T$ is the number of target nuclei per unit detector mass, $\rho_\chi$ is the density of dark matter, $m_\chi$ is the mass of DM particle, $\vec{v}_E$ is Earth's velocity relative Galaxy.
\begin{figure}[!htbp]
\includegraphics[width=\linewidth]{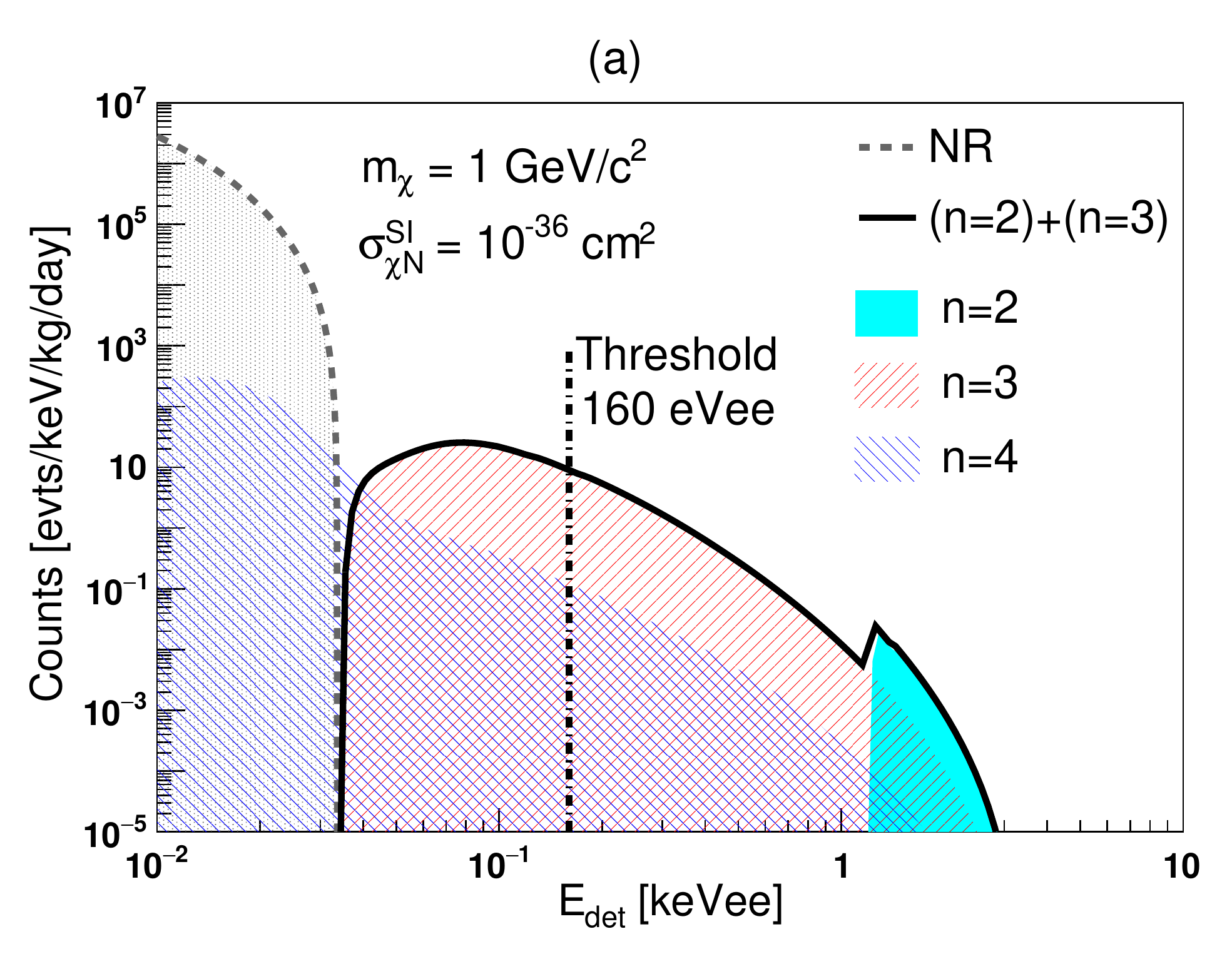}
\includegraphics[width=\linewidth]{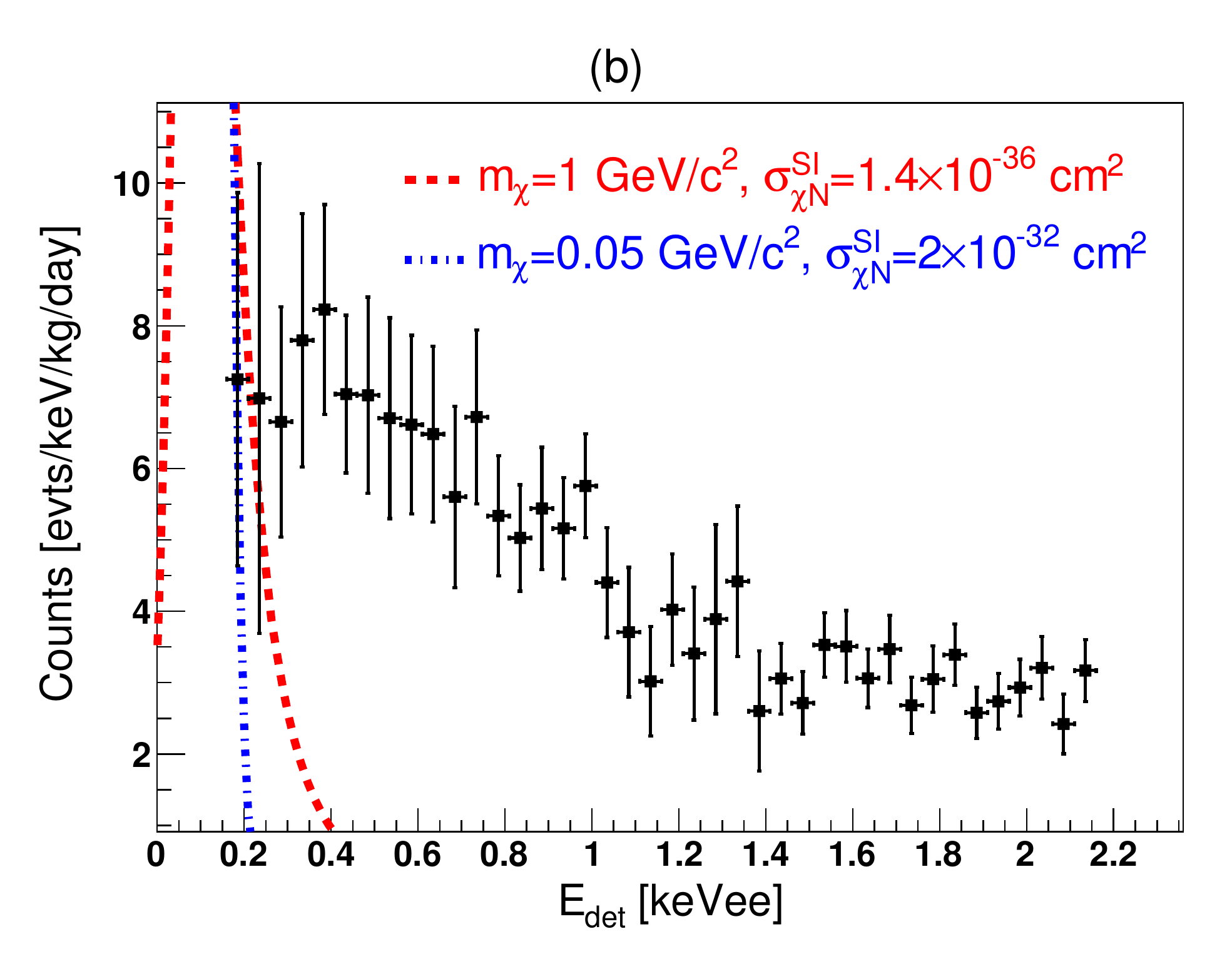}
\caption{
(a) Expected measureable spectra of the $\chi$-$N$ elastic SI-scattering (gray dash line), $\chi$-$N$ inelastic SI-scattering due to Migdal effect with N shell ($n$=4) electron, M shell ($n$=3) electron and L shell ($n$=2) electron ionized (blue, red and cyan regions, respectively), and the Migdal signal used in this analysis (($n$=2)+($n$=3), black soild line). The target nucleus is Ge, the mass of WIMPs is set to 1 GeV/$c^2$, and $\sigma_{\chi N}^{\rm SI}$ is set to $10^{-36}$ $\rm cm^2$. The analysis energy threshold is marked by the black dash-dotted line. Energy resolution is not taken into account in this plot. (b) The measured spectrum for TI analysis (black point)~\cite{cdex1b2018}, with L/M-shell x-ray contributions from the cosmogenic nuclides in the germanium crystal subtracted. The bin width is 50 eVee, and the energy range is 0.16-2.16 keVee. The blue dash-dotted line and red dash line are the expected $\chi$-$N$ spectra due to Migdal effect at $m_{\chi}$ equal to 50 MeV/$c^2$ and 1.0 GeV/$c^2$, at cross section corresponding to the upper limit at 90\% confidence level, derived by binned poisson method.
}
\label{fig::expectedspectrum}
\end{figure}

As the CDEX PPCGe detectors do not discriminate nuclear recoils from electron recoils, the observable signals are the summation of nuclear recoil energy and electron recoil energy, denoted as $E_{det} = E_{EM}+Q_{nr}E_R$, where $Q_{nr}$ is the quenching factor~\cite{soma2016}. In this letter, the Lindhard formula~\cite{lindhard} is adopted for the evaluation of $Q_{nr}$. There is no experiment data for $Q_{nr}$ of Ge below 0.2 keVnr, so the $\kappa$ value ($\kappa$=0.22) in Lindhard formula is derived by fitting of experiment data~\cite{Jones71,Jones75,cogent07} under 2 keVnr with a conservative uncertainty of 30\% adopted as systematic error.

\emph{Time-integrated (TI) analysis.}— The expected energy spectra of $\chi$-N SI scattering are shown in Fig.~\ref{fig::expectedspectrum} (a), where the target nucleus is Ge, $m_{\chi}$ = 1 GeV/$c^2$, and $\sigma_{\chi N}^{\rm SI}$ = $10^{-36}$ $\rm cm^2$. The standard WIMP galactic halo assumption and conventional astrophysical models~\cite{astropara} are used, with $\chi$-density $\rho_\chi$ set to 0.3 GeV/($c^2\rm cm^3$), Earth's velocity $v_E$ at 232 km/s, $\chi$-velocity distribution assumed to be Maxwellian with the most probable velocity $v_0=220$ km/s, the local Galactic escape velocity at 544 km/s, and the Helm form factor~\cite{formfactor1,formfactor2} is adopted. Smearing due to energy resolution is taken into account in this work. Only the ionization spectra from $L$ and $M$ shell ($n$=2, 3) are considered in this work, while those of $K$ shell ($n=1$) cannot be ionized at $m_{\chi}<$3 GeV/$c^{2}$, and those of the valence electrons ($N$ shell, $n$=4) are not reliable, as they are easily affected by the germanium band structure due to the small binding energy.

Using Fig.~\ref{fig::expectedspectrum} (a) as illustration, the expected rates in complete energy range for $\chi$-$N$ elastic SI-scattering and Migdal effects at $m_\chi$=1 GeV/$c^2$ are in ratio of about $4\times10^4:1$ (only the ionization of L and M shell electrons is considered here). However, at a threshold of 160 eVee where the $\chi$-$N$ elastic scatterings are no longer observable, the Migdal effect can still produce signals above threshold and therefore open the sensitivity windows to lower $m_\chi$.

Data used for the TI analysis are from March 2014 to July 2017, with a total exposure of 737.1 kg$\cdot$day~\cite{cdex1b2018}. The dead time ratio of the data acquisition (DAQ) system was less than 0.1\% and remained stable. 
Energy calibration is performed with internal x-ray peaks and test pulser measurements and was linear with a deviation less than 0.4\%.
The candidate $\chi$-$N$ events were selected by a series of data analysis criteria. The time coincidence of events in NaI(Tl) detector and Ge detector is used to veto the gamma-ray background events. Physical events were selected out from noise events with a combined efficiency of 17\% at the analysis threshold of 160 eVee~\cite{cdex1b2018}. Surface events were rejected and bulk events were selected based the rise time of the signal pulses. The residual energy spectrum is shown in Fig.~\ref{fig::expectedspectrum} (b).

\begin{figure}[!htbp]
\includegraphics[width=\linewidth]{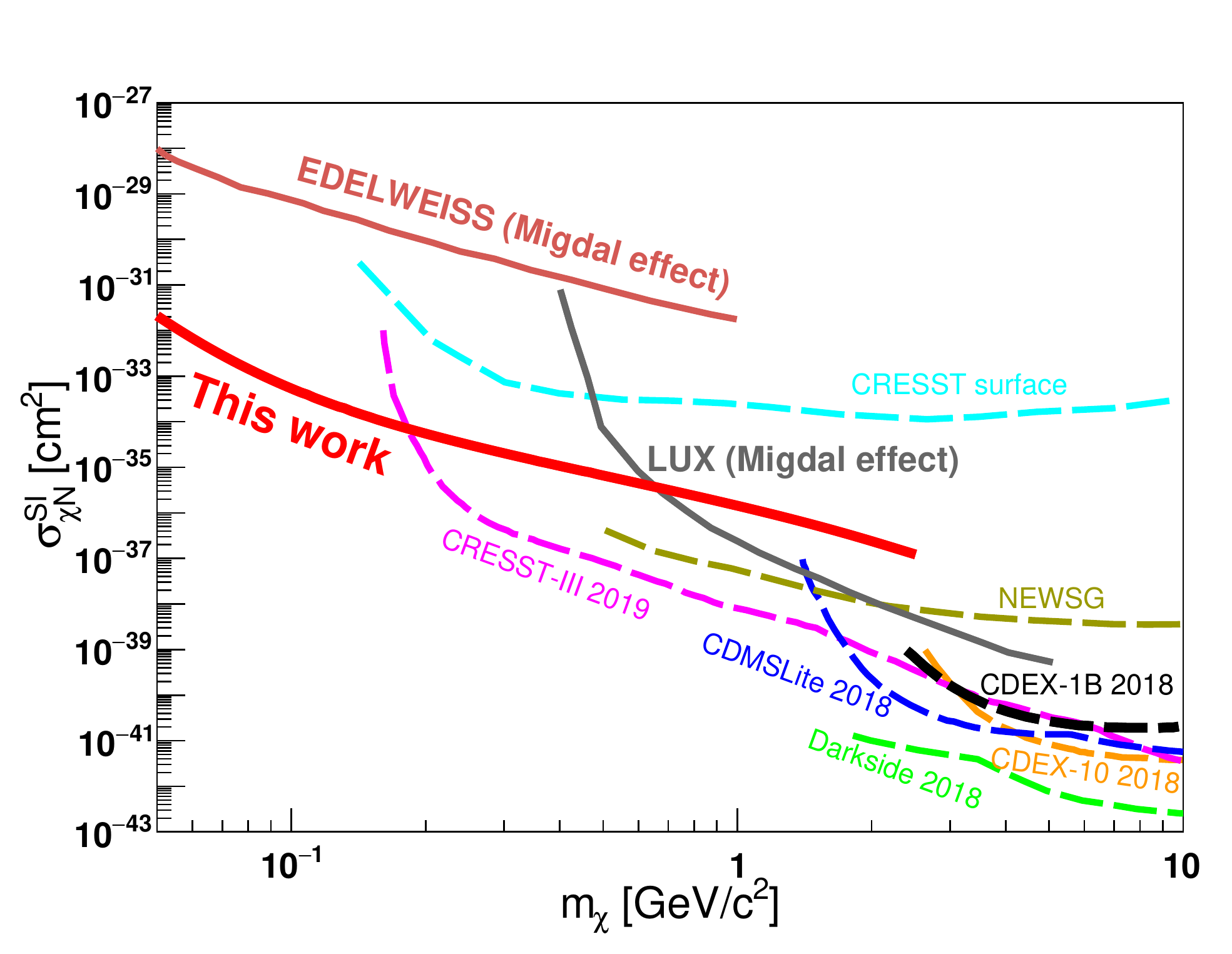}
\caption{
Upper limits at 90\% C.L. on $\sigma_{\chi N}^{\rm SI}$ derived by Binned Poisson method in TI analysis using the CDEX-1B experiment data, with several benchmark experiments~\cite{cdex1b2018,cdex102018,cresst,cdmslite,darkside,luxmigdal,cresstsurface,newsg,edelweiss} superimposed. Limits from nuclear recoil-only analysis with the same data set is shown (black dash line) as comparison. This analysis incorporating Migdal effect (red solid line) provides the best sensitivities for $m_\chi\sim$50$-$180 MeV/$c^2$, significantly expanding the excluded parameter space over earlier work (other solid lines).
}
\label{fig::exclusionSI}
\end{figure}

Upper limits at 90\% confidence level (C.L.) in $\sigma_{\chi N}^{\rm SI}$ are derived by Binned Poisson method~\cite{binpoisson}. The constraint results at $m_{\chi}$=1 GeV/$c^2$ and $m_{\chi}$=50 MeV/$c^2$ are shown in Fig.~\ref{fig::expectedspectrum} (b) by dash and dash-dotted lines. The exclusion curve is shown in Fig.~\ref{fig::exclusionSI}, in which several other experiments are superimposed for reference. New limits are achieved for $m_\chi<180$ MeV/$c^2$, and the lower reach of $m_{\chi}$ is extended to 50 MeV/$c^2$.
\begin{figure}[!htbp]
\includegraphics[width=\linewidth]{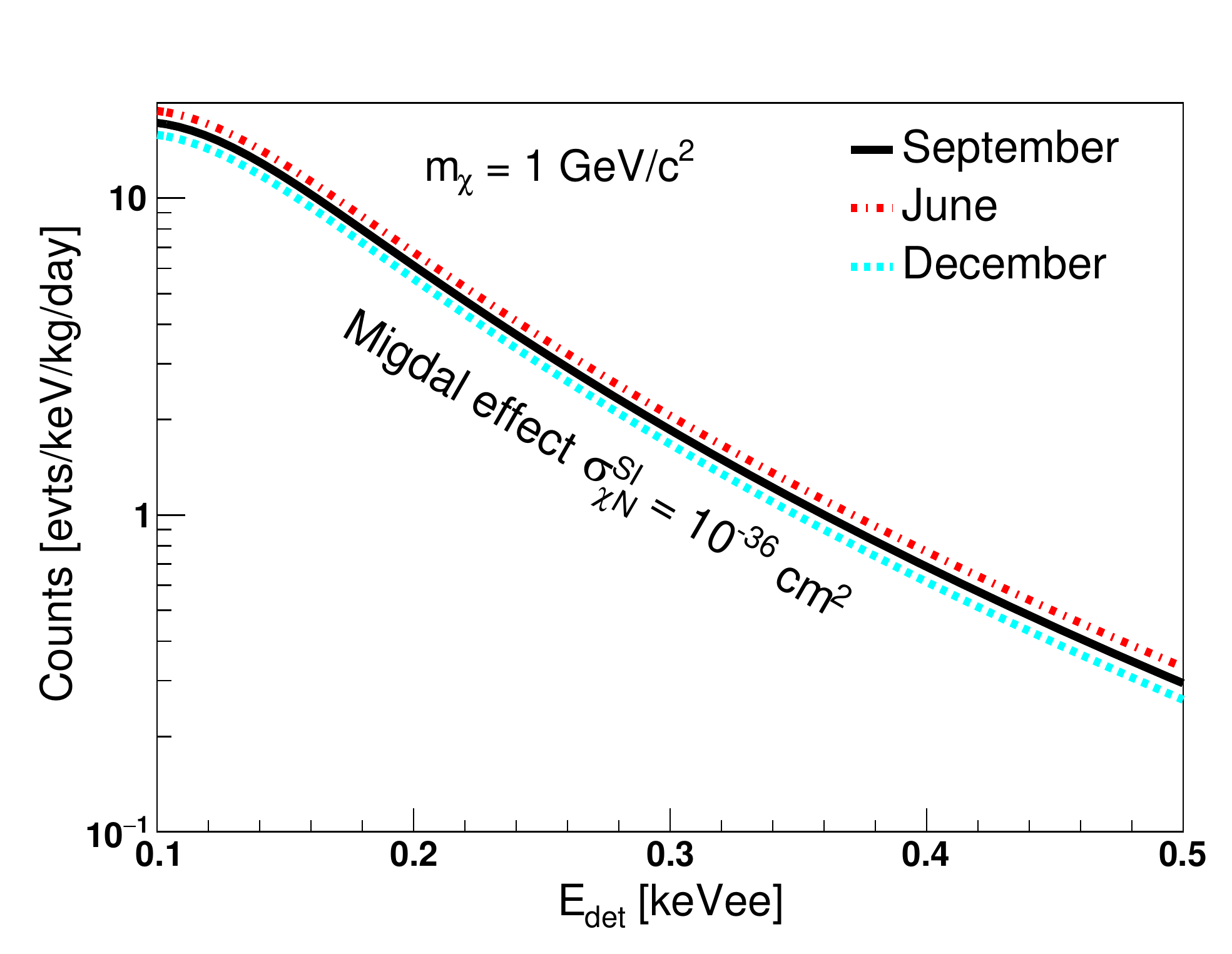}
\caption{
The curves are the expected spectra due to Migdal effect in June (red dash-dotted line), September (black solid line) and December (cyan dash line), where $m_{\chi}$ is equal to 1 GeV/$c^{2}$, $\sigma_{\chi N}^{\rm SI}$ is equal to $10^{-36}$ $\rm cm^{2}$. The energy resolution is taken into account, the standard deviation of which is 33.5 + 13.2$\times E^{\frac {1}{2}}$ (eV), where $E$ is expressed in keV.
}
\label{fig::expectedam}
\end{figure}

\emph{Annual modulation (AM) analysis.}— Positive observations of AM would provide smoking-gun signatures for WIMPs independent of the astrophysics and background models. Compared to TI analysis, the AM effects are enhanced at low WIMP mass, related to the specific shape of the ionization probability spectrum, and the sub-GeV sensitivities of the Migdal analysis can further exploit the potentials of AM studies. The Earth's velocity relative to the galactic WIMP halo is time-varying with a period of one year, and can be expressed as $v_E=232+30\times 0.51\cos(2\pi /T\times(t-\Phi))$ km/s, where $T$ is set to be 365.25 days, $\Phi$ is set to be 152.5 days from January $1^{st}$~\cite{am_theory}. The expected measurable spectra at different time of the year are shown in Fig.~\ref{fig::expectedam}, where obvious modulation effect can be observed. 

We adopt in this AM analysis the same data as previously used to study AM effects in the conventional $\chi$-$N$ nuclear recoil channel~\cite{C1B-AM-paper-underreview}. There are two datasets, Run-1 with the the NaI(Tl) anti-Compton detector, and Run-2 without NaI(Tl), having 751.3 and 428.1 live days, respectively, and together spanning a total of 1527 calendar days ($\sim$4.2 yr) and a total exposure of 1107.5 kg$\cdot$day. The background stability and environment parameters have been checked, and the time stability of the candidate $\chi$-$N$ event rates at different energy ranges were demonstrated with Fig. 1 of Ref.~\cite{C1B-AM-paper-underreview}. The Model-Dependent AM analysis~\cite{C1B-AM-paper-underreview} is adopted in this analysis. The AM-amplitudes ($A_{ik}$ of the $i$th energy bin of the $k$th Run) are related and constrained by a known function ($f$) of $m_\chi$ and $\sigma_{\chi N}^{\rm SI}$, such that $A_{ik}=\sigma_{\chi N}^{\rm SI}(m_\chi)f(E_{ik},\delta E_{ik};m_\chi)$, where $E_{ik}$ and $\delta E_{ik}$ are the mean energy and its corresponding bin-size, respectively. The same analysis threshold of 250 eVee and $\chi^2$ minimization procedures are adopted~\cite{C1B-AM-paper-underreview}. 

\begin{figure}[!htbp]
\includegraphics[width=\linewidth]{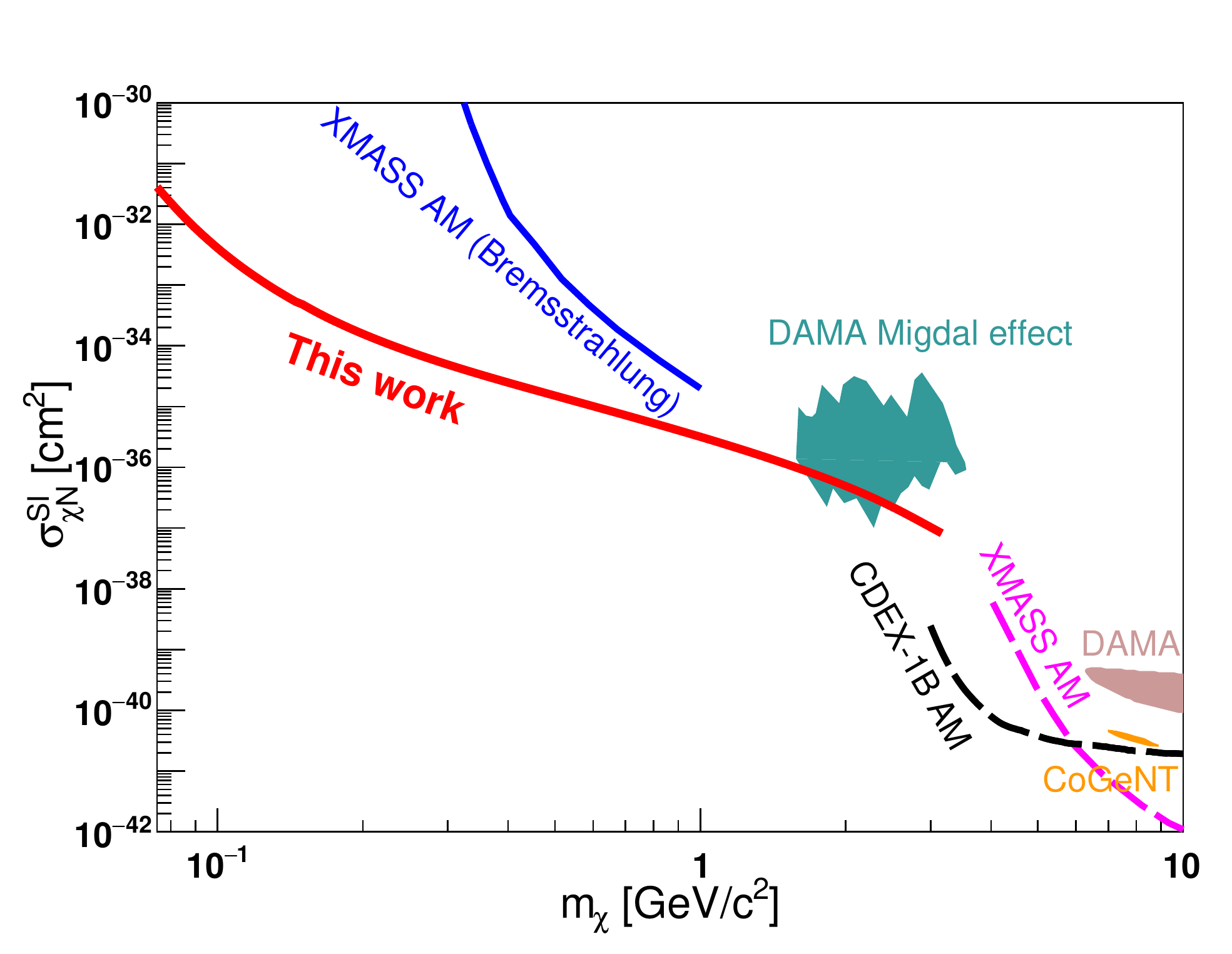}
\caption{
Upper limit at 90\% C.L. on $\sigma_{\chi N}^{\rm SI}$ derived by AM analysis using CDEX-1B data about 4.2 years~\cite{C1B-AM-paper-underreview}. The energy threshold of the modulated analysis is 250 eVee, while the reach of the exclusion line (red solid line) is extended to 75 MeV/$c^2$. The nuclear recoil only limits from the same data set is superimpsed~\cite{C1B-AM-paper-underreview}. Constraints~\cite{xmassbre} and allowed regions~\cite{cogent2013,cogent2014,dama2013,binpoisson,migdalorigin2,Baum:2018ekm} from AM analysis from other experiment are also shown.
}
\label{fig::exclusionAM}
\end{figure}

Depicted in Fig.~\ref{fig::exclusionAM} are the 90\% C.L. limits from AM analysis with Migdal effect. The only previous AM analysis at sub-GeV range was performed by XMASS-I~\cite{xmassbre} where threshold is higher (1 keVee). Its limits are also displayed. Superimposed for comparison are the AM nuclear recoils bounds from CDEX-1B~\cite{C1B-AM-paper-underreview} and XMASS-I~\cite{xmassbre}, as well as the AM-allowed regions of CoGeNT~\cite{cogent2013,cogent2014} and DAMA/LIBRA~\cite{migdalorigin2,dama2013,binpoisson,Baum:2018ekm}. The lower reach of $m_\chi$ is extended to 75 MeV/$c^2$. 

Good time stability of the CDEX-1B data leads to comparable sensitivities among the results from the AM and TI analysis due to enhanced AM effects at low $m_{\chi}$, despite the higher energy threshold of AM analysis. The measured ratios of $A_{ik}$ to the averaged background rates are less than 10\% at $E_{det}<$1 keVee for this data set. These can be compared with the expected range from Halo model and Migdal effect, which increases from $\sim$10\% to $>$40\% as $m_{\chi}$ decreases from 5 GeV/$c^2$ to 75 MeV/$c^2$.  

\emph{Summary.}— In this letter, we incorporate a newly identified mechanism on $\chi$-$N$ SI-interactions to the analysis of CDEX-1B data, based on the theoretical formula in Ref.~\cite{migdaleffect}. New $m_\chi$ windows are opened and new limits are derived. The exclusion region in $m_\chi$ can be extended down to 50 MeV/$c^2$ with an energy threshold of 160 eVee in the TI analysis. The best sensitivity in $\sigma_{\chi N}^{\rm SI}$ is achieved for $m_{\chi}\sim$(50$-$180) MeV/$c^2$ via the Migdal effect. About 4.2 years time span of CDEX-1B data are used in the AM analysis. At an energy threshold of 250 eVee, the best sensitivity of $\sigma_{\chi N}^{\rm SI}$ for $m_\chi<3.0$ GeV/$c^2$ via the Migdal effect is achieved, extending to 75 MeV/$c^2$.

For completeness, we note that bremsstrahlung effects in $\chi$-$N$ scattering were also recently derived in Ref.~\cite{bremsstrahlung} which allow light WIMP of MeV-GeV mass range to be probed. The sensitivities, however, are expected to be orders of magnitude worse than those of Midgal effects, and in the parameter space where the earth shielding effect~\cite{earthshielding,earthshielding1,earthshielding2,earthshielding3} would play a role in defining the exclusions. In addition, the excluded regions in this work have upper bounds due to earth shielding effect, detailed calculation of which is postponed to future work.

This work was supported by the National Key Research and Development Program of China (Grant No. 2017YFA0402200), the National Natural Science Foundation of China (Grants No. 11725522, No. 11675088, No.11475099) and the Tsinghua University Initiative Scientific Research Program (Grant No. 20197050007).

\bibliography{migdaleffect.bib}

\end{document}